\documentclass[aps,showpacs,prd,twocolumn,amsmath,amssymb]{revtex4-1}
\usepackage{color}
\usepackage{graphics}
\usepackage{epsfig}
\usepackage{amsmath}
\usepackage{amssymb}
\usepackage{amsthm}
\usepackage[mathscr]{eucal}
\newcommand{\slsh}[1]{{\not \! #1}}

\newcommand{\bea}{\begin{eqnarray}}
\newcommand{\eea}{\end{eqnarray}}
\newcommand{\beas}{\begin{eqnarray*}}
\newcommand{\eeas}{\end{eqnarray*}}
\newcommand{\sumint}{\sum\!\!\!\!\!\!\!\!\int}

\begin{document}

\title{Paraelectricity in Magnetized Massless QED}
\author{Efrain J. Ferrer, Vivian de la Incera and Angel Sanchez}
\affiliation{ Department of Physics, University of Texas at El Paso, 500 W. University Ave., El Paso, TX 79968, USA}

\begin{abstract}

We show that the chiral-symmetry-broken phase of massless QED in the presence of a magnetic field exhibits strong paraelectricity. A large anisotropic electric susceptibility develops in the infrared region, where most of the fermions are confined to their lowest Landau level, and dynamical mass and anomalous magnetic moment are generated via the magnetic catalysis mechanism. The nonperturbative nature of this effect is reflected in the dependence of the electric susceptibility on the fine-structure constant. The strong paraelectricity is linked to the electric dipole moments of the particle-antiparticle pairs that form the chiral condensate. The significant electric susceptibility can be used as a probe to detect the realization of the magnetic catalysis of chiral symmetry breaking in physical systems.
\end{abstract}
\maketitle

Effects of strong magnetic fields in QED have been an active research area for many years \cite{QED-B}. At present, such studies have been reactivated by the observation of very strong fields, in the range of $10^{12}-10^{16}$ G, in the surface of stellar compact objects. Also, by both theoretical and experimental indications that the colliding heavy ions at RHIC and other collaborations can generate very strong magnetic fields, estimated to be of order $eH\sim 2m^2_\pi (\sim10^{18} G)$ for the top collision, $\surd s_{NN}$ 200 GeV, in noncentral Au-Au impacts at RHIC, or even larger, $eH\sim 15 m^2_\pi (\sim 10^{19} G)$, at future LHC experiments  \cite{LHC}.

On the other hand, the study of theories of massless relativistic fermions has recently gained new interest in the context of quasiplanar systems, such as pyrolitic graphites (HOPG) \cite{Semenoff, graphite} and graphene \cite{graphene}, because their low-energy excitation
quasiparticle spectrum have a linear dispersion. The dynamics of those charge carriers is described by a "relativistic" quantum field theory of massless fermions in 2+1 dimensions ~\cite{Semenoff, Wallace}.

Massless QED in the presence of a magnetic field exhibits a peculiar phenomenology. Because of the Landau quantization of the fermion's transverse momentum in a magnetic field, the dynamics of the lowest Landau level (LLL) particles is 1+1-dimensional. This dimensional reduction favors the formation of a chiral condensate, even at the weakest coupling, because there is no energy gap between the infrared fermions in the LLL and the antiparticles in the Dirac sea. This phenomenon is known as the magnetic catalysis of chiral symmetry
breaking (MC$\chi$SB).  The MC$\chi$SB modifies the vacuum properties and induces dynamical parameters that depend on the applied field. This effect has been actively investigated for the last two decades~\cite{severalaspects}-\cite{ferrerincera}.
In the original studies of the MC$\chi$SB ~\cite{severalaspects}-\cite{leungwang}, the catalyzed chiral condensate was assumed to give rise only to a dynamical fermion mass. Recently, however, it has become clear \cite{ferrerincera} that besides the dynamically generated mass, the MC$\chi$SB inevitably produces also a dynamical anomalous magnetic moment (AMM), because this second parameter does not break any symmetry that has not already been broken by the chiral condensate and the magnetic field. The dynamical AMM leads, in turn, to a
nonperturbative Lande g-factor and Bohr magneton proportional to the inverse of the
dynamical mass. The induction of the AMM leads to a nonperturbative Zeeman effect \cite{ferrerincera}.

An important aspect of the MC$\chi$SB is its universal character. It will occur in any relativistic theory of interactive massless fermions in a magnetic field. The MC$\chi$SB has been proposed as the mechanism explaining various effects in quasiplanar condensed matter systems~\cite{MC-Applications}.

A drawback of the MC$\chi$SB phenomenon is that the dynamical parameters (mass and AMM) are extremely small even at relatively high fields. Aside from the fact that it may be experimentally challenging measuring the magnetically catalyzed parameters, there may be cases where other competing mechanisms are proposed to explain a given magnetic field effect. Consequently, it would be nice to have an independent way to experimentally distinguish the MC$\chi$SB from other possibilities. One of the main purposes of this letter is to argue that by using a weak electric field as a probe, one could obtain, by measuring the induced electric polarization, compelling evidence in favor or against the existence of MC$\chi$SB.

The electric polarization is found as minus the derivative of the electromagnetic free energy with respect to the applied electric field. For a weak electric field $E$, the free-energy density can be expanded in powers of $E$ as
\begin{equation}\label{Free-energy}
\Phi=\Phi_0-\eta E-\chi E^2+...
\end{equation}
In a magnetized medium, the coefficients $\Phi_0$, $\eta$, $\chi$, etc., may in principle depend on the magnetic field. The susceptibility coefficient $\eta$ is different from zero for ferroelectric materials \cite{ferroelectricity}. In magnetized QED it is zero, because the second term in the r.h.s. of (\ref{Free-energy}) violates parity, a symmetry that is not broken neither in massive QED nor in the chirally broken phase of massless QED. The coefficient $\chi$ characterizes the lowest order of the system dielectric response. It accounts for the electric polarization $P=\chi E$ induced by the applied electric field. This term does not break any additional symmetry in (\ref{Free-energy}).

The electric susceptibility $\chi$ in massive QED at strong magnetic field can be obtained from the one-loop photon polarization operator in an external field configuration defined by the gauge $A_\mu=(-Ex_3,0,-Hx_1,0)$, which implies the existence of uniform parallel magnetic and electric fields in the $x_3$-direction. Since the electric field is only playing the role of a probe, while we are looking for a strong magnetic field effect, we take the approximation, $|eE|<m^2<|eH|$, where $m$ is the electron mass. Hence, in the calculation of the photon polarization operator the fermion Green's functions depend only on the magnetic field, while the effect of the weak electric field can be neglected. Thus, the general covariant structure of the polarization operator under such conditions is \cite{Shabad-1}
\bea
 \Pi_{\mu\nu}(q)=\sum_{a=1}^3\kappa_a\frac{b_\mu^{(a)}b_\nu^{(a)}}{(b^{(a)})^2},
\label{Pi-covariant}
\eea
where $b_\mu^{(a)}$ are the orthogonal vectors, transverse to $q_\mu$,
\begin{eqnarray}\label{Orthogonal}
b_\mu^{(1)}=-q^2\widehat{F}_{\mu\nu}\widehat{F}_{\nu\rho}q_{\rho}+q_{\bot}^2q_{\mu},\qquad  \nonumber
\\
b_\mu^{(2)}=\frac{1}{2}\varepsilon_{\mu\nu\rho\lambda}\widehat{F}_{\nu\rho}q_\lambda,\quad b_\mu^{(3)}=\widehat{F}_{\mu\nu}q_\nu,
\end{eqnarray}
with $\widehat{F}_{\mu\nu}=F_{\mu\nu}/H$ denoting the normalized
electromagnetic strength tensor. In (\ref{Pi-covariant}), $\kappa_a$
are scalar coefficients depending on the magnetic field. At strong magnetic field, where the electrons will all lie in the LLL, only the coefficient $\kappa_2$ is different from zero. Thus, from (\ref{Pi-covariant}) and (\ref{Orthogonal}) we see that only the longitudinal components of $\Pi_{\mu\nu}$, which in turn depend only on the longitudinal momenta, contribute in the strong-field limit. In the static limit, $q_0=0$, and at small spatial momenta (for a constant and uniform electric field, the contribution to the free energy of powers of momentum higher than quadratic is zero, so we do not need to consider them),
the polarization-operator coefficient $\kappa_2$ behaves as \cite{Shabad-1}-\cite{coefficient-LLL}
\begin{equation}\label{Infra-red}
   \kappa_2(q_0=0,|\overrightarrow{q}|\rightarrow 0)
     \simeq \frac{-\alpha |eH|}{3\pi m^2}q_3^2
\end{equation}

In this limit, the only component different from zero is $\Pi_{00}=\kappa_2$. Its contribution to the electromagnetic free-energy density is given by
\begin{eqnarray}\label{Strong-B-Suscept}
   \Phi-\Phi_0\sim\frac{1}{V}\int A_0(x_3)\Pi_{00}(x_3-x_3')A_0(x_3')dx_3dx_3'=\nonumber\\
    =-\chi_{QED}E^2,\qquad \qquad \qquad \qquad \qquad \quad
\end{eqnarray}
with $\chi_{QED}=\alpha |eH|/3\pi m^2$,
and $V$ the system volume. Clearly, high field values $|eH| > m^2\sim 10^{13}$ G are required for the electric susceptibility ($\chi_{QED}$) to be significant. Therefore, this polarization effect can only be relevant for the astrophysics of compact stars and for heavy ion collisions, where such large field strengths can exist. It is worthy noticing, on the other hand, that the result (\ref{Infra-red}) is not valid for $m=0$. One can check that if the electron mass is taken to zero and the calculations are repeated in the strong-field limit, the electric susceptibility becomes zero. In this case, we know that the situation cannot be changed by any higher order perturbative contribution, since the chiral symmetry of the massless theory is protected against perturbative corrections. Chiral symmetry, however, can be broken nonperturbatively via $MC\chi SB$.

Our goal now is to find the electric susceptibility in the chirally broken phase of massless QED in the presence of  uniform electric and magnetic fields along the $x_3$-direction. As already discussed, along with the induced dynamical mass, the chiral condensate necessarily produces a
dynamical AMM \cite{ferrerincera}. Here again, we are looking for a strong magnetic field effect and treat the electric field as a weak probe. Hence, we assume $|eE| < (E^{0})^{2}  <|eH|$, with $E^{0}$ the dynamical LLL rest energy. Accordingly, we neglect $E$ in the fermion propagator. The one-loop photon polarization operator is
\bea
    \Pi_{\mu\nu}(x,y)
        &=&-4\pi i\alpha Tr\left[
           \gamma_\mu G(x,y)\gamma_\nu G(y,x)
            \right]
\label{oppol1}
\eea
where the electron full propagator is given by
\bea
    G(x,x')=\sumint\frac{d^4p}{(2\pi)^4}
           \mathbb{E}_p(x)\Pi(l)
           \tilde{G}^l(\overline{p})
           \overline{\mathbb{E}}_p(x'),
\label{oppol2}
\eea
with $\sumint\frac{d^4p}{(2\pi)^4}\equiv\sum_l\int\frac{dp_0dp_2dp_3}{(2\pi)^4}$, $\Pi(l)=\Delta(+)+I(1-\delta^{0l})$ \cite{leungwang}, and $l=0,1,2,...$ the LL numbers. We assume $sgn(eB)=+$. In (\ref{oppol2}) we used the $\mathbb{E}_p$ Ritus' transformation $\mathbb{E}_p(x)=\sum_{\sigma=\pm1}E_{p\sigma}(x)\Delta(\sigma)$ (originally developed for fermions in \cite{Ritus:1978cj} and later extended to vector fields in \cite{efi-ext}), where $ \Delta(\sigma)=\frac{I+i\sigma \gamma^1\gamma^2}{2},\quad \sigma=\pm 1$,
are the spin projectors, and $E_{p\sigma}(x)=N_n e^{i(p_0x^0+p_2x^2+p_3x^3)}D_n(\rho)$,
with $D_n(\rho)$ denoting the parabolic cylinder functions with argument
$\rho=\sqrt{2|eH|}(x_1-p_2/eH)$, normalization factor $N_n=(4\pi
eH)^{1/4}/\sqrt{n!}$,  and  positive integer index $n=n(l,\sigma)\equiv l+\frac{\sigma+1}{2}$.
Eq. (\ref{oppol2}) can be used to obtain the fermion propagator in momentum space as a function of the dynamical
mass $M^{(l)}$ and the magnetic energy $T^{l}$ associated to the AMM of each LL \cite{ferrerincera},
\begin{equation}\label{full-FP}
G^{l}(p,p')=(2\pi)^4\widehat{\delta}^{(4)}(p-p')\Pi(l)\widetilde{G}^{l}(\overline{p}),
\end{equation}
with
\bea
  \tilde{G}^l(\overline{p})&=&
    \sum_{\sigma,\overline{\sigma}=\pm 1}
    \frac{N^l(\sigma T,\overline{\sigma}V_{||})
          -iV_\perp^l(\Lambda_\perp^+-\Lambda^-_\perp)}
         {D^l(\sigma\overline{\sigma}T)}
    \Delta(\sigma)\Lambda^{\overline{\sigma}}_{||},
\nonumber \\
\label{ad12}
\eea
and
\begin{eqnarray}  \label{ad13}
\Lambda^{\overline{\sigma}}_{||}=
         \frac{1}{2}
         \left(1+{\overline{\sigma}}\frac{\slsh{\overline{p}}_{||}}{|\overline{p}_{||}|}
         \right), \quad
           \Lambda^\sigma_{\perp}=
         \frac{1}{2}
         \left(1+i\sigma\gamma^2
         \right),\quad \nonumber \\
N^l(\sigma T,\overline{\sigma}V_{||})
     =\sigma T^l-M^l-\overline{\sigma}V_{||}^l,\quad \quad\nonumber \\
     D^l(\sigma {\overline{\sigma}}T)=(M^l)^2-(V_{||}^l-\sigma{\overline{\sigma}} T^l)^2+(V_\perp^l)^2,\quad \nonumber \\
     V_{||}^l=(1-Z_{||}^l)|\overline{p}_{||}|,\quad V_{\perp}^l=(1-Z_\perp^l)|\overline{p}_{\perp}|\qquad
\end{eqnarray}
Here, $\overline{p}_{\|}=(p_0,0,0,p_3)$ and $\overline{p}_{\bot}=(0,0,\sqrt{2eBl},0)$ denote the longitudinal and transverse momenta, respectively, while $Z_{||}^l$ and $Z_\perp^l$ are the wave-function renormalization coefficients.
Taking into account that the MC$\chi$SB is an infrared phenomenon where the magnetic field becomes the leading parameter, we have that the main contribution to the polarization operator comes from the low-energy region where only fermions in the LLL
contribute. Thus, transforming (\ref{oppol1}) to momentum space and keeping the leading contribution $l=0$, we obtain
\bea
     \Pi^{||}_{\mu\nu}(q)&=&
      -2i\alpha |eH|
       e^{-\frac{{q^2_\perp}}{2|eH|}}
       \int \frac{d^2p}{(2\pi)^2}
\nonumber \\
     &\times&
       Tr\left[
           \gamma^{||}_\mu
           \Delta(+)\tilde{G}^{0}(\overline{p})
           \gamma^{||}_\nu
           \Delta(+)\tilde{G}^{0}(\overline{p-q})
       \right],
\label{oppol63}
\eea
with
\begin{eqnarray}\label{Pgorrito-LLL}
\widetilde{G}^{0}(\overline{p})
=\frac{\Delta(+)\Lambda^{+}_{\|}}{|\overline{p}_{||}|-(M^0+T^0)}
-\frac{\Delta(+)\Lambda^{-}_{\|}}{|\overline{p}_{||}|+(M^0+T^0)}
\nonumber
\\
+\frac{\Delta(-)\Lambda^{+}_{\|}}{|\overline{p}_{||}|-(M^0-T^0)}
-\frac{\Delta(-)\Lambda^{-}_{\|}}{|\overline{p}_{||}|+(M^0-T^0)}
\end{eqnarray}
Note that only the longitudinal components of $\Pi_{\mu\nu}$ survive in (\ref{oppol63}). Integrating in momenta, using Feynman
parametrization, and dimensional regularization yields
\begin{equation}\label{Pi-LLL-covariant}
   \Pi_{\mu\nu}^{\|}(q)=
     \overline{\kappa}_2\frac{b_\mu^{(2)}b_\nu^{(2)}}{(b^{(2)})^2},
\end{equation}
where
\bea
   &&\hspace{-0.7cm}\overline{\kappa}_2\equiv -\frac{2\alpha |eH|}{\pi}e^{-\frac{{q^2_\perp}}{2|eH|}}
\nonumber \\
&&\hspace{-0.5cm}\times
  \left[
    1+\frac{2{E^0}^2 }{q_{||} \sqrt{q^2_{||}-4 {E^0}^2}}
    \ln \left(\frac{\sqrt{q^2_{||}-4{E^0}^2}+q_{||}}
                    {\sqrt{q^2_{||}-4{E^0}^2}-q_{||}}\right)
   \right],
\label{oppol69a}
\eea
with $E^0=M^0+T^0$, the LLL rest energy.
Notice that the polarization operator (\ref{Pi-LLL-covariant}) is transverse,
$q^{\mu}\Pi_{\mu\nu}=\Pi_{\mu\nu}q^{\nu}=0$, ensuring the
gauge invariance of the LLL approximation.
In the static limit, $q_0=0$, and infrared region
$|\overrightarrow{q}|\rightarrow 0$, the coefficient $\overline{\kappa}_2$ behaves as
\begin{equation}\label{Infrared}
   \overline{\kappa}_2(q_0=0,q_3\rightarrow 0)
     \simeq \frac{-\alpha |eH|}{3\pi(E^0)^2}q_3^2
\end{equation}
This result indicates that the inclusion of the AMM term does not contribute to produce Debye screening in the infrared region. That is, at distances $r>1/E^0$, a charge within this medium interacts through a normal Coulomb potential. Comparing (\ref{Infrared}) with (\ref{Infra-red}) we see that the induced rest energy $E^{0}$ plays the same role in the broken phase of massless QED as the electron mass in massive QED. The difference is, however, that $E^{0}$ is not a fixed parameter, but it has to be found as the solution of the Schwinger-Dyson equation for the electron self-energy.

To find $E^{0}$ we follow the results of Ref. \cite{ferrerincera}, where $E^{0}$ was found as the solution of the SD equation for the electron self-energy in the ladder approximation
\bea
   \Sigma(x,x')=ie^2\gamma^\mu G(x,x')\gamma^\nu D_{\mu\nu}(x-x'),
\label{auto1}
\eea
where $\Sigma(x,x')$ is the fermion self-energy operator,
$D_{\mu\nu}(x-x')$ is the bare photon propagator in the Feynman's gauge~\cite{WT}, and $G(x,x')$ is the full fermion propagator (\ref{oppol2}). In the LLL, the SD equation reduces to \cite{ferrerincera}
\begin{equation}\label{relations-3}
1=e^2(4eH)\int\frac{d^4\widehat{q}}{(2\pi)^4}
\frac{e^{-\widehat{q}^2_\bot}}{\widehat{q}^2}\frac{1}{(E^{0})^2+
q_{\|}^2}
\end{equation}
where we introduced the normalized-momentum notation $\widehat{q}^2=q^2/2|eH|$. The solution of (\ref{relations-3}) is
given by
\begin{equation}
\label{Mass-Eq-Solution} E^{0}\simeq \sqrt{2|eH|}
\exp{-\sqrt{\frac{\pi}{\alpha}}}
\end{equation}

The electric susceptibility can be found now similarly to the QED case done previously. From (\ref{Pi-LLL-covariant})-(\ref{Infrared}) and (\ref{Mass-Eq-Solution}) we have that the medium behaves as a linear, homogeneous, and anisotropic dielectric, with electric susceptibility in the $x_{3}$-direction
\begin{equation}\label{Susc-MC}
\chi_{MC\chi SB}=\frac{\alpha |eH|}{3\pi (E^{0})^2}=\frac{\alpha}{6\pi}\exp \sqrt{\frac{4\pi}{\alpha}}
\end{equation}
 Eq.(\ref{Susc-MC}) shows that the susceptibility depends nonperturbatively on the fine-structure constant and its value is independent of the applied magnetic field. Notice the marked difference with the QCD situation at strong magnetic field ($|eH|\gg\Lambda_{QCD}$), since in QCD the MC$\chi$SB leads to a \textit{chromo}-susceptibility that remains a function of the magnetic field through the running of the strong coupling $\alpha_s$ \cite{Igor}. The colossal susceptibility (\ref{Susc-MC}) characterizes the electric response of the system to an electric field parallel to the magnetic one. At zero temperature, no critical magnetic field strength is required to catalyze the chiral symmetry breaking. On the other hand, even though the magnitude of the electric polarization does not depend on the magnetic field, for a fixed weak electric field, it must satisfies the condition $P< \frac{\alpha|H|}{3\pi^{2}}$. That is, increasing $H$ increases $E^{0}$, and a larger electric field probe, constrained by the condition $|eE| < (E^{0})^{2}  <|eH|$, is allowed. We should underline that from a physical point of view, it is natural to expect a large anisotropic electric susceptibility in the phenomenon of MC$\chi$SB in QED, as the ground state of the system is characterized by pairs forming tiny electric dipoles that can be polarized by the external electric field. The role of the magnetic field here is to induce the pairs, while the role of the electric field is to polarize them.

The dramatic increase of the electric susceptibility produced by the magnetically catalyzed chiral pairs can be the best candidate to probe whether the $MC\chi SB$ mechanism is taking place or not. In a system of massless fermions in a magnetic field, if the system exhibits a sizable electric polarization under the application of a weak electric field probe along the direction of the magnetic field, it will be a plausible evidence of the $MC\chi SB$ phenomenon.

An important implication of this result is that the chirally broken phase exhibits strong paraelectricity, a property found in certain condensed matter systems like quantum paraelectric (QP) materials ~\cite{Paraelectricity} and transition metal oxides (TMO) \cite{TMO}. In those materials, unaligned electric dipoles are aligned in an external electric field, producing a high electric susceptibility, often exceeding $10^4$. In QP materials the large electric susceptibility is temperature-independent below certain critical temperature, a property attributed to a quantum phase transition~\cite{Paraelectricity}. An interesting question to explore in the future is whether the strong susceptibility found here within a (3+1)-dimensional theory is also present in quasiplanar condensed matter systems as bilayer graphene. It is known, that the band structure of bilayer graphene can be controlled by an applied electric field perpendicular to the layers' plane. The electric field creates an electronic gap between the valence and conduction bands with energy values that varies from zero to midinfrared~\cite{Castro}, depending on the field strength. Under a very weak electric field the gap is practically zero and the spectrum is Dirac-like. Even though this is a very peculiar 3+1-D system, only formed by two layers, one could attempt to model it with a 3+1-D theory of interactive massless fermions. Because of the universality of the MC$\chi$SB, we expect that the application of a strong magnetic field parallel to the weak electric one will trigger the generation of a dynamical energy gap. Under these conditions, one would expect that detecting a very large electric susceptibility in the direction of the applied fields would signal the realization of the MC$\chi$SB mechanism.

{\bf Acknowledgments:} This work has been supported in part by the DOE Nuclear Theory grant DE-SC0002179.

\end{document}